\documentclass[showpacs,showkeys,preprint,aps]{revtex4}

\usepackage{graphicx}
\usepackage{dcolumn}
\usepackage{bm}

\newcommand{\ppbar}{$p\overline{p}$}
\newcommand{\ccbar}{$c\overline{c}$}
\newcommand{\bbbar}{$b\overline{b}$}
\newcommand{\ttbar}{$t\overline{t}$}
\newcommand{\MET}{\mbox{$\raisebox{.3ex}{$\not$}E_T$}}
\newcommand{\gevcc}{GeV/$c^{2}$}
\newcommand{\isofo}{$ISO_{\Delta R = 0.4}$}
\newcommand{\isose}{$ISO_{\Delta R = 0.7}$}
\newcommand{\chargino}{$\tilde{\chi}_1^{\pm}$}
\newcommand{\neutralino}{$\tilde{\chi}_2^0$}
\newcommand{\zvert}{$z_{vertex}$}
\newcommand{\ctcres}{$\delta p_T/p_T^2 = 8\times10^{-4}$ (GeV/$c$)$^{-1}$}
\newcommand{\calres}{$13.5\%/\sqrt{E \sin \theta} \oplus 2\%$}
\newcommand{\squark}{$\tilde{q}$}
\newcommand{\gluino}{$\tilde{g}$}

\font\eightit=cmti8
\def\r#1{$^{#1}$}

\begin{document}


\begin{large}
\begin{center}
Inclusive Search for Anomalous Production of High-$p_T$ 
Like-Sign Lepton Pairs in \ppbar~Collisions at $\sqrt{s} 
= 1.8$ TeV
\end{center}
\end{large}

\font\eightit=cmti8
\def\r#1{\ignorespaces $^{#1}$}
\hfilneg
\begin{sloppypar}
\noindent
D.~Acosta,\r {14} T.~Affolder,\r 7 M.G.~Albrow,\r {13} 
D.~Ambrose,\r {36} D.~Amidei,\r {27} K.~Anikeev,\r {26} 
J.~Antos,\r 1 G.~Apollinari,\r {13} T.~Arisawa,\r {50} 
A.~Artikov,\r {11}
% ...
\end{sloppypar}
\vskip .026in
\begin{center}
(CDF Collaboration)
\end{center}
\vskip .026in
\begin{center}
\r 1 {\eightit Institute of Physics, Academia Sinica, 
Taipei, Taiwan 11529, Republic of China} \\ 
\r 2 {\eightit Argonne National Laboratory, 
Argonne, Illinois 60439} \\  
\r 3 {\eightit Istituto Nazionale di Fisica Nucleare, 
University of Bologna, I-40127 Bologna, Italy} \\  
\r 4 {\eightit Brandeis University, Waltham, Massachusetts 
02254} \\  
\r 5 {\eightit University of California at Davis, 
Davis, California  95616} \\  
\r 6 {\eightit University of California at Los Angeles, 
Los Angeles, California  90024} \\   
\r 7 {\eightit University of California at Santa Barbara, 
Santa Barbara, California 93106} \\   
\r 8 {\eightit Instituto de Fisica de Cantabria, 
CSIC-University of Cantabria, 39005 Santander, 
Spain} \\  
\r 9  {\eightit Carnegie Mellon University, 
Pittsburgh, Pennsylvania  15213} \\  
\r {10} {\eightit Enrico Fermi Institute, 
University of Chicago, Chicago, Illinois 60637} \\  
\r {11} {\eightit Joint Institute for Nuclear Research, 
RU-141980 Dubna, Russia} \\  
\r {12} {\eightit Duke University, Durham, North Carolina 
27708} \\  
\r {13} {\eightit Fermi National Accelerator Laboratory, 
Batavia, Illinois 60510} \\  
\r {14} {\eightit University of Florida, Gainesville, Florida 
32611} \\  
\r {15} {\eightit Laboratori Nazionali di Frascati, 
Istituto Nazionale di Fisica Nucleare, I-00044 Frascati, 
Italy} \\  
\r {16} {\eightit University of Geneva, CH-1211 Geneva 4, 
Switzerland} \\  
\r {17} {\eightit Glasgow University, Glasgow G12 8QQ, United 
Kingdom}\\  
\r {18} {\eightit Harvard University, Cambridge, Massachusetts 
02138} \\  
\r {19} {\eightit Hiroshima University, Higashi-Hiroshima 724, 
Japan} \\  
\r {20} {\eightit University of Illinois, Urbana, Illinois 
61801} \\  
\r {21} {\eightit The Johns Hopkins University, 
Baltimore, Maryland 21218} \\  
\r {22} {\eightit Institut f\"{u}r Experimentelle Kernphysik, 
Universit\"{a}t Karlsruhe, 76128 Karlsruhe, Germany} \\  
\r {23} {\eightit Center for High Energy Physics: Kyungpook National 
University, Taegu 702-701; Seoul National University, Seoul 151-742; 
and SungKyunKwan University, Suwon 440-746; Korea} \\  
\r {24} {\eightit Ernest Orlando Lawrence Berkeley National Laboratory, 
Berkeley, California 94720} \\  
\r {25} {\eightit University College London, London WC1E 6BT, 
United Kingdom} \\  
\r {26} {\eightit Massachusetts Institute of Technology, Cambridge,
Massachusetts  02139} \\     
\r {27} {\eightit University of Michigan, Ann Arbor, Michigan 
48109} \\  
\r {28} {\eightit Michigan State University, 
East Lansing, Michigan  48824} \\  
\r {29} {\eightit Institution for Theoretical and Experimental Physics, 
ITEP, Moscow 117259, Russia} \\  
\r {30} {\eightit University of New Mexico, 
Albuquerque, New Mexico 87131} \\  
\r {31} {\eightit Northwestern University, Evanston, 
Illinois  60208} \\  
\r {32} {\eightit The Ohio State University, Columbus, Ohio  
43210} \\  
\r {33} {\eightit Osaka City University, Osaka 588, 
Japan} \\  
\r {34} {\eightit University of Oxford, Oxford OX1 3RH, United 
Kingdom} \\  
\r {35} {\eightit Universita di Padova, Istituto Nazionale di Fisica 
Nucleare, Sezione di Padova, I-35131 Padova, Italy} \\  
\r {36} {\eightit University of Pennsylvania, 
Philadelphia, Pennsylvania 19104} \\     
\r {37} {\eightit Istituto Nazionale di Fisica Nucleare, University 
and Scuola Normale Superiore of Pisa, I-56100 Pisa, 
Italy} \\  
\r {38} {\eightit University of Pittsburgh, 
Pittsburgh, Pennsylvania 15260} \\  
\r {39} {\eightit Purdue University, West Lafayette, Indiana 
47907} \\  
\r {40} {\eightit University of Rochester, Rochester, New York 
14627} \\  
\r {41} {\eightit Rockefeller University, New York, New York 
10021} \\  
\r {42} {\eightit Instituto Nazionale de Fisica Nucleare, 
Sezione di Roma, University di Roma I, ``La Sapienza," 
I-00185 Roma, Italy} \\  
\r {43} {\eightit Rutgers University, Piscataway, New Jersey 
08855} \\  
\r {44} {\eightit Texas A\&M University, College Station, Texas 
77843} \\  
\r {45} {\eightit Texas Tech University, Lubbock, Texas 
79409} \\  
\r {46} {\eightit Institute of Particle Physics, University of Toronto, 
Toronto M5S 1A7, Canada} \\  
\r {47} {\eightit Istituto Nazionale di Fisica Nucleare, 
University of Trieste/Udine, Italy} \\  
\r {48} {\eightit University of Tsukuba, Tsukuba, Ibaraki 305, 
Japan} \\  
\r {49} {\eightit Tufts University, Medford, Massachusetts 
02155} \\  
\r {50} {\eightit Waseda University, Tokyo 169, Japan} \\  
\r {51} {\eightit University of Wisconsin, Madison, Wisconsin 
53706} \\  
\r {52} {\eightit Yale University, New Haven, Connecticut 06520}
\end{center}

\date{\today}

\begin{abstract}
We report on a search for anomalous production of events with at 
least two charged, isolated, like-sign leptons with $p_T > 11$ 
GeV/$c$ using a 107 pb$^{-1}$ sample of 1.8 TeV \ppbar~collisions 
collected by the CDF detector.  We define a signal region containing 
low background from Standard Model processes.  To avoid bias, we fix 
the final cuts before examining the event yield in the signal 
region using control regions to test the Monte Carlo predictions.  
We observe no events in the signal region, consistent with an 
expectation of $0.63^{+0.84}_{-0.07}$ events.  We present 95\% 
confidence level limits on new physics processes in both a 
signature-based context as well as within a representative minimal 
supergravity ($\tan\beta = 3$) model.
\end{abstract}


\maketitle

Numerous attempts to resolve theoretical problems with the standard 
model (SM) require the existence of new particles with masses at the 
electroweak scale, $\sim$~100~\gevcc~\cite{leptoquark,axion,susy}.  
A productive method of searching for new particles at this scale 
has been to search for lepton production at \ppbar~colliders with 
high momentum transverse to the beam axis ($p_T$).  Such searches 
led to the discovery of the $W$ and $Z$ bosons at the 
CERN $S$\ppbar$S$ and to the discovery of the $t$ quark at the 
Fermilab Tevatron.  Recently, searches for anomalous high-$p_T$ 
lepton production have been used to constrain supersymmetric 
extensions to the SM.  For example, production of charginos 
($\tilde{\chi}^{\pm}$) and neutralinos ($\tilde{\chi}^0$) were 
constrained by searches for events with three high-$p_T$ 
leptons~\cite{trilep-cdf,trilep-d0}.  Limits on gluino 
production were likewise placed by searching for events 
with two like-sign leptons, two jets, and transverse energy 
imbalance, \MET~\cite{cdf-gluino,d0-gluino}.  These analyses 
achieved the necessary suppression of background processes by 
requiring three or more reconstructed objects in the final state 
and constraints on their kinematical properties.

In this Letter we present a search for new particles
with masses at the electroweak scale using a minimal number of
required objects or kinematical cuts.  Specifically, we search 
for two like-sign, isolated leptons in the final state, 
but do not require any other objects or \MET.  We define a signal 
region with less than one event expected from SM background but 
broad acceptance for typical models of new particle production 
resulting in like-sign signatures~\cite{ssd,maj-nu,dc-higgs}.  To 
avoid bias, we fix the final cuts before examining the event yield 
in the signal region~\cite{blind}.

We examine 107 pb$^{-1}$ of data collected by the Collider Detector 
at Fermilab (CDF) during the 1992-95 data run of the Tevatron.  
The CDF detector~\cite{cdfdetector} is an azimuthally and
forward-backward symmetric solenoidal detector designed to 
study \ppbar~reactions at the Tevatron.  A time projection 
chamber measures the distance of the \ppbar~collision event vertex 
(\zvert) from the center of the detector along the beam direction.  
The central tracking chamber (CTC) measures the trajectories of 
charged particles traversing a uniform 1.4 T magnetic field with 
a resolution of \ctcres.  Outside the solenoid, a lead/scintillator 
central electromagnetic sampling calorimeter detects electromagnetic 
showers with an energy resolution of \calres. Steel/scintillator 
hadronic calorimeters directly behind the electromagnetic 
calorimeters measure the hadronic component of deposited energy.  
Drift chambers located behind the steel detect muon candidates with 
momenta above 3 GeV/$c$.

We begin with a sample of 457,478 loosely selected dilepton 
events~\cite{donethesis}.  We select candidate events in a manner 
similar to previous CDF trilepton searches~\cite{trilep-cdf}.  
We identify charged leptons as electrons or muons each with $p_T > 11$ 
GeV/$c$ using the ``strict'' selection criteria of those analyses.  
As in the previous analyses, we remove events consistent with 
photon conversions or cosmic rays.  We reject background where 
one lepton is a partner of known resonances by removing events 
consistent with any $\psi$, $\Upsilon(1S)$, or $Z$ resonance.  
We require $|z_{vertex}| <$ 60 cm and $|z_{lepton} - z_{vertex}| <$ 
5 cm for each lepton, to ensure that both leptons came from the same 
primary collision and are well measured.  We identify the two 
highest-$p_T$ leptons with like-sign charges as the like-sign (LS) 
dilepton pair.  To reduce background from back-to-back QCD dijet 
events in which both jets are misidentified as leptons, we require 
the lepton pair to have vector sum transverse momentum 
of $p_T^{\ell\ell}~>$~20~GeV/$c$ and invariant mass 
of $m_{\ell\ell}~>$~10~\gevcc.

One notable difference from previous analyses is a modification 
to the lepton isolation variable ($ISO$) which separates leptons 
from jets.  $ISO$ is the scalar sum of the transverse energy 
($E_T$) measured in each calorimeter cell, $\sum E_T$, added in 
quadrature to the scalar sum of the $p_T$ measured in the CTC, 
$\sum p_T$, within a cone 
$\Delta R~\equiv~\sqrt{(\Delta\phi)^2~+~(\Delta\eta)^2}~=~0.4$ 
of each lepton candidate.  The energy of the lepton candidate is 
removed from the $ISO$ sum by subtracting the $p_T$ of the lepton 
candidate track, $p_T^{cand}$, and the calorimeter $E_T$ of the 
lepton candidate, $E_T^{cand}$, from $\sum p_T$ and $\sum E_T$, 
respectively,
\begin{equation}
ISO = \sqrt{ ( \sum E_T - E_T^{cand} )^2 + ( \sum p_T - p_T^{cand} )^2 }.
\end{equation}
\noindent $E_T^{cand}$ is the scalar sum of the $E_T$ in the 
calorimeter cell to which we extrapolate the lepton candidate track 
(the ``seed'' cell) and the two cells adjacent to either side of 
the seed cell in $\eta$ to account for lateral energy leakage 
between cells: $E_T^{cand}~=~E_T^{seed}~+~E_T^{leakage}$.
In this analysis we have changed $E_T^{leakage}$ so that 
the lepton is excised from $\sum E_T$ more effectively by modeling 
the energy leakage between cells in greater detail~\cite{mattsthesis}.  
In addition to the usual cut at \isofo~$<$~2~GeV we have added a cut 
on an identically defined $ISO$ cone with radius of 
0.7, \isose~$<$~7~GeV.  The combination of double-cone cut and 
redefined $E_T^{leakage}$ is the ``new'' $ISO$.

\begin{figure}
\includegraphics[scale=0.7]{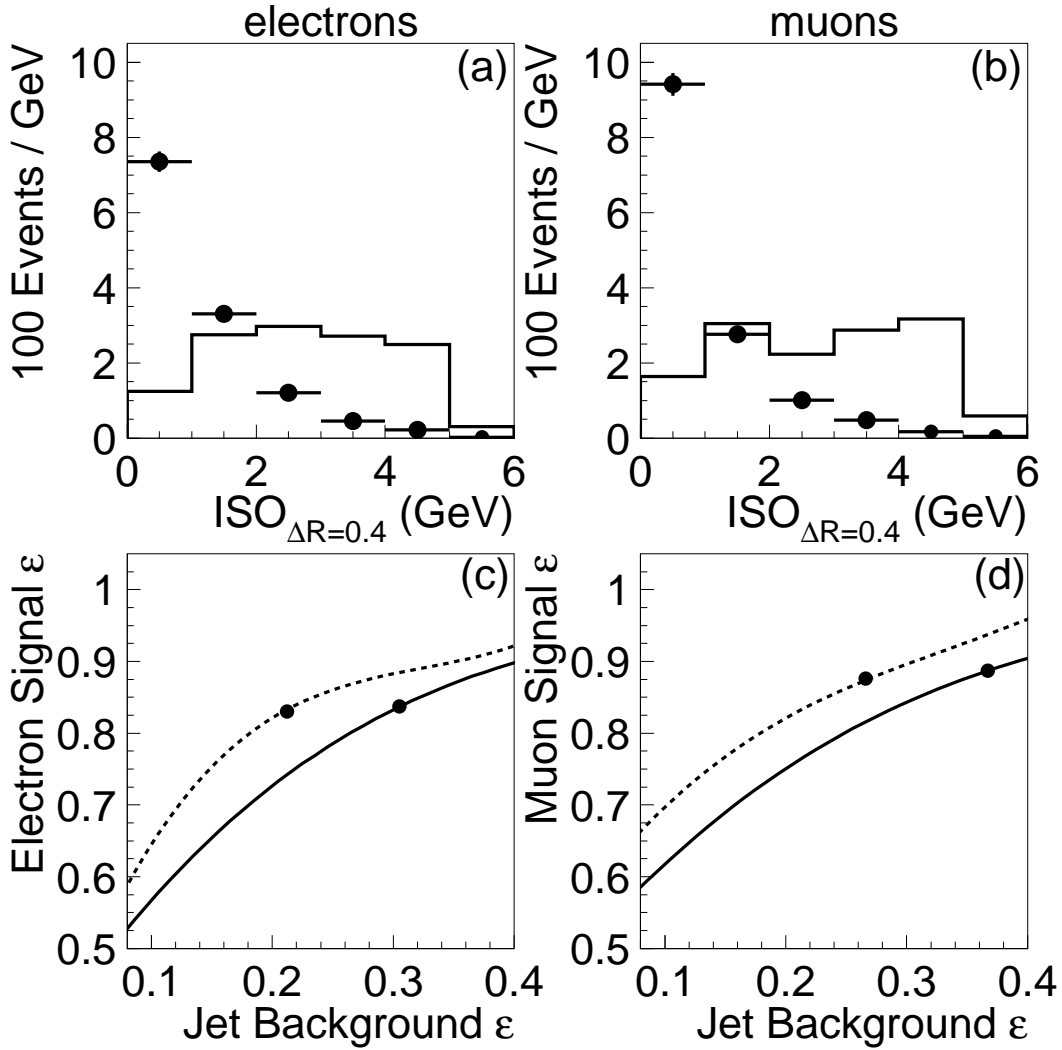}
\caption{\label{isoplot} (a) and (b) show the energy in the 
\isofo~cone for electrons and muons, respectively, in the $Z$ 
dilepton dataset (points) and jet background dataset (histogram).  
Errors shown are statistical only.  (c) and (d) show $\epsilon$ 
for electrons and muons, respectively, as a function of jet 
background $\epsilon$ for original $ISO$ (solid line) and 
new $ISO$ (dashed line).  The markers indicate the position of 
the nominal original and new $ISO$ cuts on the $\epsilon$ curves.}
\end{figure}

To evaluate the efficacy of the new $ISO$ cut we demonstrate an 
increased separation of lepton signal from background from jets 
misidentified as leptons with the new $ISO$.  We select leptons 
from 1255 $Z\rightarrow e^+ e^-$ and 1389 $Z\rightarrow \mu^+ \mu^-$ 
events with no $ISO$ requirement on the leptons.  From bias-removed 
jet control samples with 20 and 50 GeV thresholds on the jet 
$E_T$ we select a sample of 292 (237) jets passing all the LS 
dilepton analysis electron (muon) identification requirements 
except $ISO$.  Figures~\ref{isoplot}(a) and (b) compare the energy 
in the new \isofo~cone between the lepton and jet samples for 
electrons and muons, respectively.  Figures~\ref{isoplot}(c) and (d) 
show efficiency ($\epsilon$) for electrons and muons, 
respectively, from the $Z\rightarrow \ell^+ \ell^-$ samples as a 
function of $\epsilon$ for background from the jet control samples 
for original and new $ISO$.  We generate the $\epsilon$ curves by 
varying the \isofo~cut between 1 and 4 GeV within each sample.  
With the nominal cuts, new $ISO$ reduces background from jets 
being misidentified as leptons by a factor of two from the 
original cut while retaining the same efficiency for leptons.

Diboson production, $WZ$ and $ZZ$, where ``$Z$'' denotes a mixture 
of the $Z$ and $\gamma^*$, produces an irreducible source of SM 
background.  Although the $Z$ resonance cut removes most of these 
events, some survive because the $Z$ is off-shell or we fail to 
find one of the leptons from the $Z$.  We model this background 
using the Monte Carlo programs {\small{PYTHIA}}~\cite{pythia} and 
{\small{MCFM}}~\cite{mcfm} which include off-shell contributions.  
The two processes contribute $0.25\pm 0.09$ and $0.07\pm 0.02$ 
events, respectively, to the signal region.  The only other 
significant background is $W$+jets and $Z$+jets production where 
one of the jets is misidentified as a lepton.  Because the rate 
of lepton misidentification is beyond the scope of the Monte Carlo 
programs and simulations, we anchor this calculation in the data.  
First, we verify that {\small{PYTHIA}} correctly models the observed 
rate of isolated tracks as a function of $p_T$ 
in $Z\rightarrow \ell^+ \ell^-$ events, excluding the two tracks 
from the legs of the $Z$.  Second, we use several control samples 
to measure the probability that such isolated tracks pass all 
lepton ID requirements: $(2.5\pm 0.7)$\% with no measurable $p_T$ 
dependence.  Third, we multiply the {\small{PYTHIA}} prediction for 
production of a $W$ or $Z$ with an underlying isolated track by 
this factor to estimate backgrounds to be $0.30\pm 0.08$ 
and $0.03\pm 0.01$ events, respectively; for a complete description 
of this method see~\cite{backup}.  Using {\small{ISAJET}}~\cite{isabug}
we estimate the small contribution, $0.008^{+0.006}_{-0.004}$, 
from \ttbar, \bbbar, and \ccbar.  Finally, we set an upper limit 
to the contribution from events in which both lepton candidates 
are jets misidentified as leptons, $0.0^{+0.83}_{-0.0}$, using 
event yields outside the signal region.  We find negligible 
background due to charge misassignment by using $Z$ events and 
track curvature studies.

\begin{figure}
\includegraphics[scale=0.7]{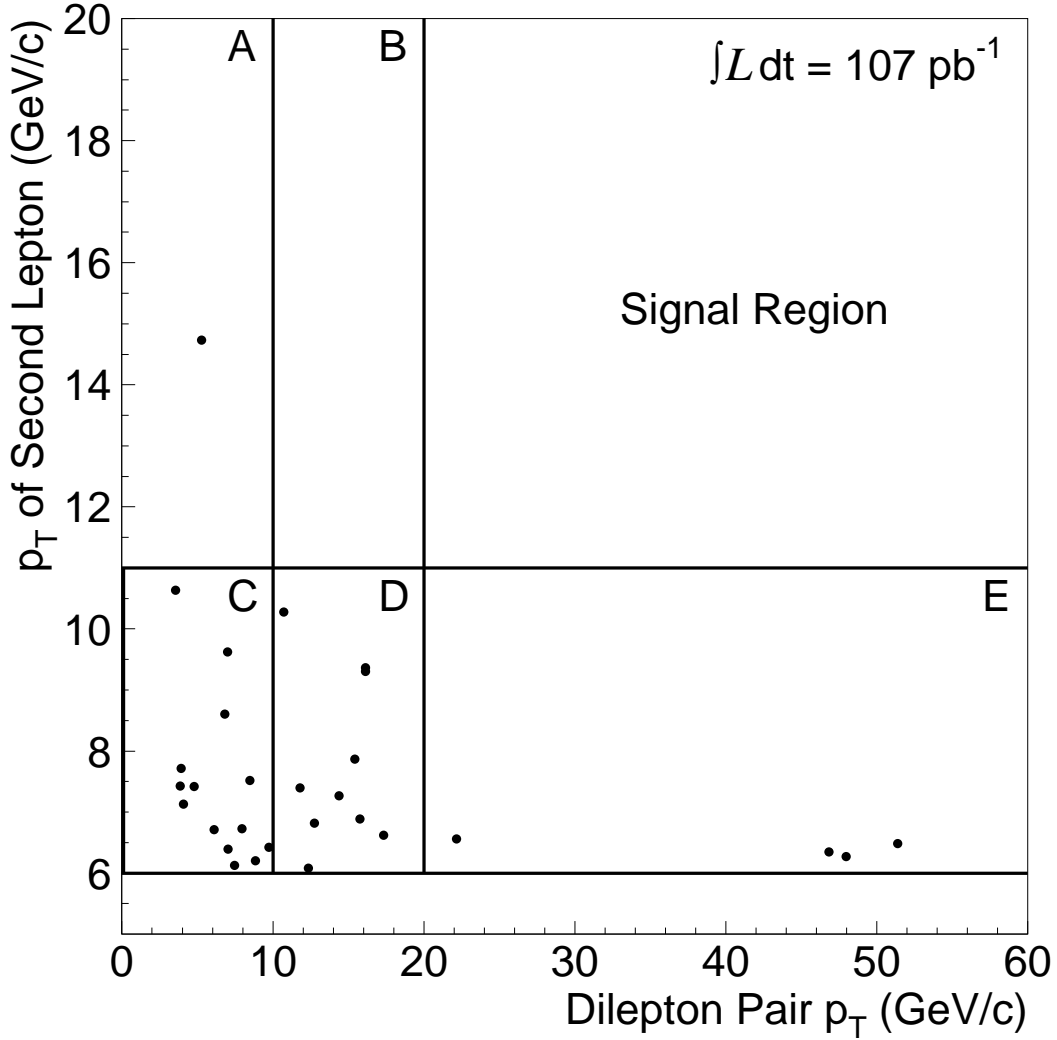}
\caption{\label{pt2vsptll} Observed events in kinematical regions 
adjacent to the signal region; see Table~\ref{jet_bkg}.}
\end{figure}

We use kinematical regions having sensitivity to different 
background sources to test the background predictions.  Events 
near the signal region shown in Figure~\ref{pt2vsptll} are 
compared to background predictions in Table~\ref{jet_bkg}.  
The consistency of these control regions indicates the 
reliability of the lepton misidentification estimates.  With 
all the nominal cuts but the $Z$-removal cut inverted, we 
predict $0.11\pm 0.03$ events and see zero.  If, instead of 
requiring like-sign, we require an opposite-sign pair and an 
additional isolated $p_T >$ 3 GeV/$c$ track we predict $68\pm 9$ 
events and observe 62, thereby testing the Monte Carlo modeling 
of the effect of lost Drell-Yan leptons.

\begin{table}
\caption{\label{jet_bkg} Comparison of SM background to events 
selected in the data in the control regions shown in 
Figure~\ref{pt2vsptll}.}
\begin{ruledtabular}
\begin{tabular}{cccc}
Region & Background(s) & Expected background & Data \\
\hline
A & QCD dijet & $2.2^{+1.8}_{-1.5}$ & 1 \\
B & $WZ$,$ZZ$ & $0.1^{+0.9}_{-0.1}$ & 0 \\
C & QCD dijet & $19.7\pm8.4$ & 14 \\
D & QCD dijet & $10.0\pm4.5$ & 10 \\
E & $W$+jets, QCD dijet & $6.0^{+1.6}_{-1.3}$ & 4 \\
\end{tabular}
\end{ruledtabular}
\end{table}

In the signal region we predict $0.63^{+0.84}_{-0.07}$ total events 
and 
observe zero.  Hence this analysis provides no indication of physics 
beyond the SM and we proceed to set limits on new physics using a 
Bayesian technique~\cite{bayes}.  Following the methodology of 
previous analyses, we apply sources of systematic uncertainty 
including trigger efficiency, luminosity, lepton ID efficiency, 
structure function choice, and $Q^2$~variations~\cite{trilep-cdf} to 
each model of particle production considered below.

\begin{figure}[b]
\includegraphics[scale=0.7]{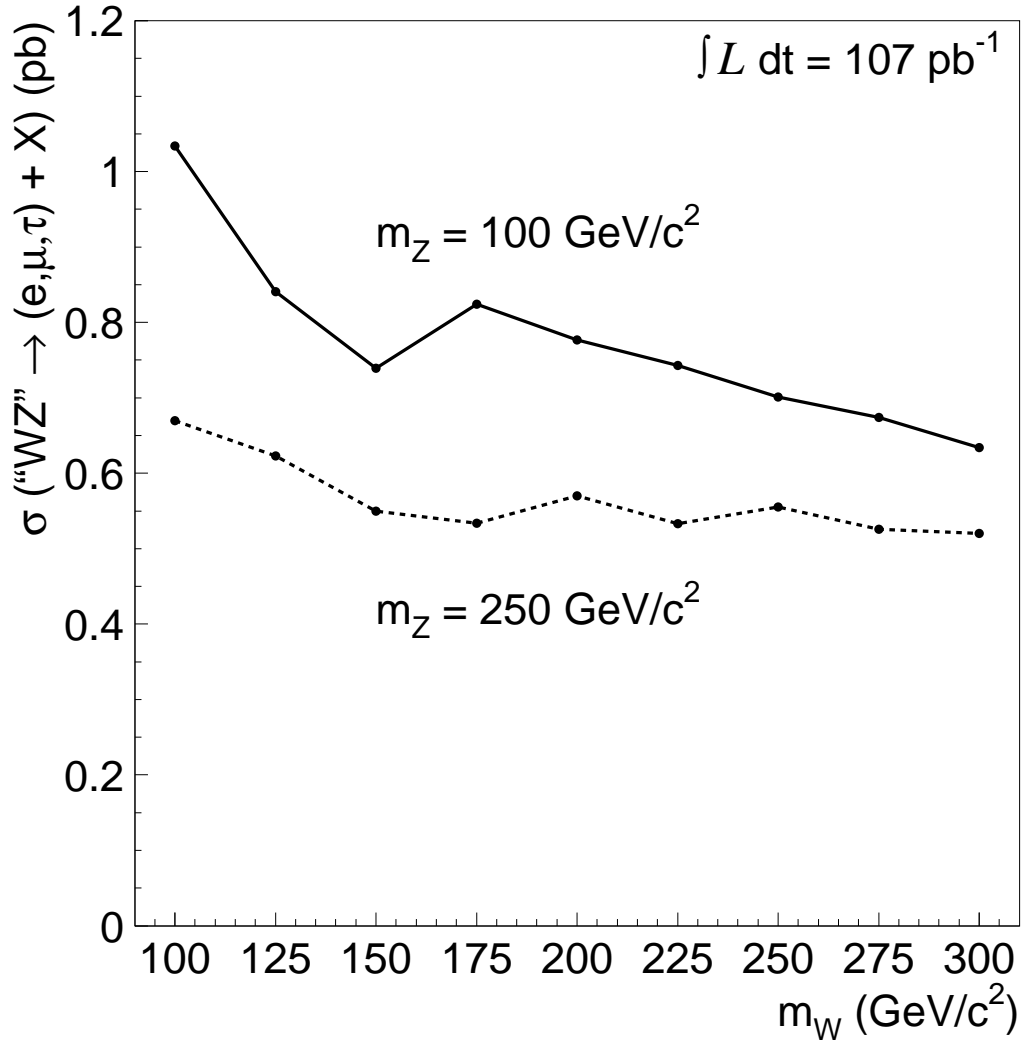}
\caption{\label{gp_lim_plot} The 95\% confidence level limit on 
the cross section for ``$WZ$-like'' production as a function of 
the $W$-like particle mass, for two representative masses of 
the $Z$-like particle.}
\end{figure}

Because we perform this search without considering any one particular 
model for new physics, we evaluate the result as a general limit on 
particle production leading to the LS dilepton signature.  As an 
example, we generate $WZ$ pairs with {\small{PYTHIA}} using standard 
couplings and spins.  However, we allow the masses of the $W$-like 
and $Z$-like particles to vary.  After forcing the bosons to decay 
leptonically, we find the efficiencies range from 3\% to 8\% as the 
$W$-like and $Z$-like masses vary from 100 to 300 \gevcc.  Exclusion 
limits on the cross section times branching ratio including a 16\% 
systematic uncertainty are shown in Figure~\ref{gp_lim_plot}.

\begin{figure}
\includegraphics[scale=0.7]{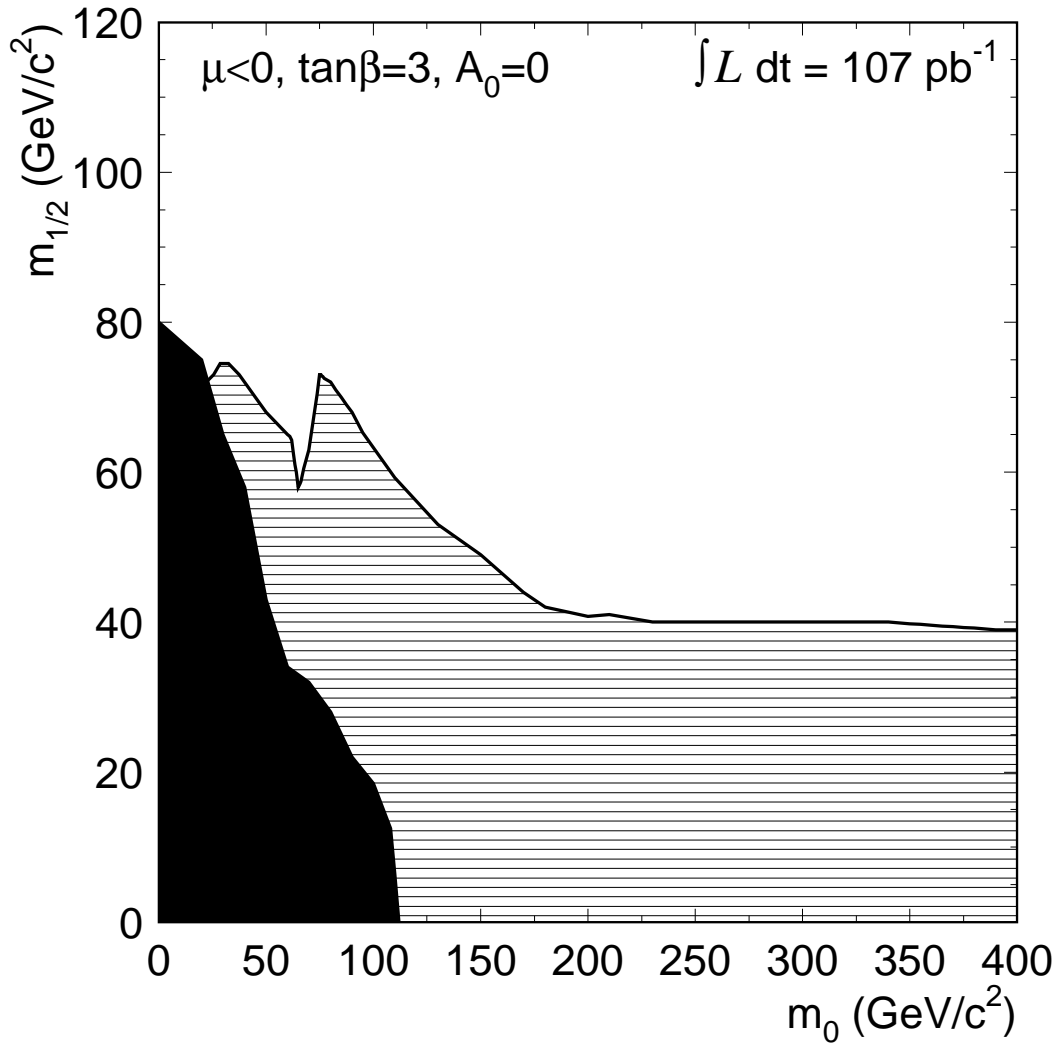}
\caption{\label{m0mhalflimit} The 95\% confidence level limit on the 
parameters $m_0$ and $m_{1/2}$ in the mSUGRA framework 
for $\tan\beta$~=~3, $\mu~<$~0 and $A_0$~=~0 (hatched region).  
The shaded region is theoretically excluded.  The dip near 
75~\gevcc~results from the loss of sensitivity to the 
\chargino\neutralino~signal due to decays of \chargino~and 
\neutralino~to sneutrinos.  At lower $m_0$, the limit is regained 
due to sensitivity to \squark~and \gluino~production~\cite{msugra}.}
\end{figure}

In addition to such signature-based limits we derive a limit within 
the framework of mSUGRA~\cite{msugra}, a supergravity-inspired 
extension to the Minimal Supersymmetric Standard Model~\cite{susy}.  
We take representative parameters $\tan\beta$ = 3, $\mu <$ 0 and 
$A_0 =$ 0, but allow $m_0$ and $m_{1/2}$ to vary and use {\small{PYTHIA}} 
to calculate event yields.  The simulation allows all particles to 
decay according to their calculated branching ratios so that charged 
leptons may be produced at any stage of cascade sparticle and 
particle decays.  Within the context of this model, the selection 
is reoptimized according to the 95\% confidence expected upper 
limit on the signal cross section, leading to an improved 
sensitivity by lowering the $p_T^{\ell \ell}$ cut 
from 20 to 10~\gevcc.  In this mSUGRA model LS dilepton events are 
primarily produced by the decay \chargino\neutralino $\rightarrow$ 
$\ell^{\pm}\ell^{\pm}\ell^{\mp}\tilde{\chi}_1^0\tilde{\chi}_1^0 \nu$.  
However, this analysis is sensitive as well to LS dileptons produced 
in the sequential decays of squarks (\squark) and gluinos (\gluino) 
and even to production of \chargino~with \gluino.  Here the 
efficiency, which includes the branching ratio to leptons imposed 
by the model, ranges from 0.02\% to 0.12\%.  We calculate exclusion 
limits on the cross section as a function of $m_0$ and $m_{1/2}$, 
including a 17\% systematic uncertainty, to construct an excluded 
region in $m_0-m_{1/2}$ space, as shown in Figure~\ref{m0mhalflimit}.  
We use the available next-to-leading order corrections (20\%--40\%) 
to the cross sections~\cite{prospino,plehn}.  Previous exclusions, 
based on \MET~in multijet events~\cite{multijets}, have already 
covered all of this space, but with an entirely different technique.

We have shown in feasibility studies~\cite{backup} that the LS 
dilepton signature considered here and the previously published 
trilepton signatures~\cite{trilep-cdf} can be significantly 
complementary.  The Run II data can be analyzed simultaneously with 
both techniques to obtain a sensitivity to mSUGRA space greater than 
either analysis alone.

We thank the Fermilab staff and the technical staffs of the 
participating institutions for their vital contributions.  This 
work was supported by the U.S. Department of Energy and National 
Science Foundation; the Italian Istituto Nazionale di Fisica 
Nucleare; the Ministry of Education, Culture, Sports, Science, 
and Technology of Japan; the Natural Sciences and Engineering 
Research Council of Canada; the National Science Council of the 
Republic of China; the Swiss National Science Foundation; the A.P. 
Sloan Foundation; the Bundesministerium fuer Bildung and Forchung, 
Germany; and the Korea Science and Engineering Foundation (KoSEF); 
the Korea Research Foundation; and the Comision Interministerial 
de Ciencia y Tecnologia, Spain.

\bibliography{lsdprl_apspreprint}

\end{document}